\documentclass[preprint]{elsarticle}
\usepackage{graphicx}
\usepackage{amssymb,amsmath}
\usepackage{subfig}
\usepackage{color}
\newcommand{\nc}{\newcommand}
\nc{\rnc}{\renewcommand}
\nc{\x}[1]{\mbox{#1}}           
\nc{\hs}[1]{\hspace*{#1}} \nc{\vs}[1]{\vspace*{#1}}
\rnc{\theta}{\vartheta} \rnc{\rho}{\varrho}
\rnc{\epsilon}{\varepsilon} \nc{\preload}{\Delta} \nc{\scal}{\kappa}
\nc{\red}[1]{\textcolor{red}{#1}} \nc{\nh}{N_{\mathrm{h}}}
\nc{\dd}{{\mathrm{d}}} \nc{\herm}{^{\mathrm H}}
\nc{\ii}{{\mathrm{i}}} \nc{\ee}{{\mathrm{e}}}
\nc{\mm}[1]{\mathbf{#1}}
\nc{\mms}[1]{\boldsymbol{#1}}
\nc{\suml}[2]{\sum \limits_{#1}^{#2}} \nc{\intl}[2]{\int
\limits_{#1}^{#2}} \rnc{\matrix}[2]{\left[\!\!\begin{array}{#1}
#2\end{array}\!\!\right]} \rnc{\vector}[1]{\matrix{c}{#1}}
\nc{\inv}{^{-1}} \nc{\tra}{^{\mathrm T}} \nc{\sgn}{\mathrm{sgn}}
\nc{\kn}{{k_{\mathrm n}}} \nc{\kt}{{k_{\mathrm t}}}
\nc{\fc}{F_{\mathrm R}}
\newcommand{\moptc}{m_{\mathrm{opt,c}}}
\nc{\ommod}{\omega_0}
\nc{\uudist}{\,\,\,}
\nc{\fnl}{\mm g}
\nc{\reliab}{RA}
\nc{\dmod}{D}
\nc{\amod}{q}
\nc{\stim}{\epsilon}
\nc{\areference}{a_{\mathrm{ref}}}
\nc{\prob}{\operatorname{Pr}}
\nc{\perf}{\Theta}
\nc{\phaslag}{\theta}
\nc{\perfref}{\Theta_{\mathrm{ref}}}
\nc{\fst}{F_0}
\nc{\nnn}{[1]}
\nc{\ttt}{[2]}
\nc{\ntd}{N_{\mathrm{td}}}
\rnc{\Re}[1]{\operatorname{Re}\left\{ #1 \right\}}
\rnc{\Im}[1]{\operatorname{Im}\left\{ #1 \right\}}
\nc{\kreis}[1]{\mbox{\mbox{\text{\Large{\ensuremath{\bigcirc}}}}\hspace{-2.4ex}#1\hspace{1.2ex}}}
\nc{\prog}[1]{{\sf{#1}}}
\nc{\matlab}{\prog{Matlab}}
\nc{\ie}{i.\,e.\,}
\nc{\eg}{e.\,g.\,}
\nc{\cf}{cf.\,}
\nc{\etal}{et~al.\,}
\nc{\z}[2]{\x{\sc #1}~{\x{\cite{#2}}}}
\nc{\zo}[1]{\x{\cite{#1}}}
\nc{\fabstand}{\,}
\nc{\fp}{\fabstand.}
\nc{\fk}{\fabstand,}
\nc{\tab}[5][tbh]{\begin{table}[#1]\centering\caption{#4\label{tab:#5}}\begin{tabular}{#2}\hline #3 \\ \hline\end{tabular}\end{table}}
\newcommand{\fss}[4][tbh]{%
 \begin{figure}[#1]
 \centering
 \if@draft
  \framebox[150mm]{\raisebox{0mm}[25mm][25mm]{\texttt{#2}}}
 \else
  \includegraphics[width=#4\textwidth]{Figures/#2}
 \fi
 \caption{#3}
 \label{fig:#2}
\end{figure}
}

\nc{\e}[2]{\begin{equation} #1 \label {eq:#2} \end{equation}}
\nc{\ea}[2]{
\begin{eqnarray}
#1 \label {eq:#2} \end{eqnarray}}
\nc{\eal}[3][0.0ex]{
\begin{samepage}
\begin{eqnarray*}
#2
\end{eqnarray*}
\nopagebreak[4] \vs{#1} \nopagebreak[4]\vs{-2ex} \nopagebreak[4]
\begin{eqnarray}
\label {eq:#3}
\end{eqnarray}
\end{samepage}\hs{-0.35em}}
\nc{\g}[1]{{$#1$}}
\nc{\fref}[1]{{Fig.~\ref{fig:#1}}}
\nc{\frefo}[1]{{\ref{fig:#1}}}
\nc{\frefoo}[1]{{#1}}
\nc{\frefs}[1]{{Figs.~\ref{fig:#1}}}
\nc{\tref}[1]{{Tab.~\ref{tab:#1}}}
\nc{\trefo}[1]{{\ref{tab:#1}}}
\nc{\trefs}[1]{{Tab.~\ref{tab:#1}}}
\nc{\erefn}[1]{{Eq.~(\ref{#1})}}
\nc{\eref}[1]{{Eq.~(\ref{eq:#1})}} \nc{\erefo}[1]{(\ref{eq:#1})}
\nc{\erefs}[2]{{Eqs.~(\ref{eq:#1}) and (\ref{eq:#2})}}
\nc{\sref}[1]{{Section~\ref{sec:#1}}}
\nc{\srefo}[1]{\ref{sec:#1}}
\nc{\srefs}[1]{{Sections~\ref{sec:#1}}}
\nc{\aref}[1]{{{Appendix~\ref{asec:#1}}}}
\nc{\arefo}[1]{{\ref{asec:#1}}}
\nc{\arefs}[1]{{{Appendices~\ref{asec:#1}}}}
\nc{\ssref}[1]{{Subsection~\ref{sec:#1}}}
\nc{\ssrefo}[1]{\ref{sec:#1}}
\nc{\ssrefs}[1]{{Subsections~\ref{sec:#1}}}

\begin{document}

\begin{frontmatter}

\title{Reliability Optimization of Friction-Damped Systems Using
Nonlinear Modes}

\author[ids]{Malte Krack\corref{cor1}}
\ead{krack@ila.uni-stuttgart.de}
\author[ids]{Sebastian Tatzko}
\author[ids]{Lars Panning-von Scheidt}
\author[ids]{J\"org Wallaschek}


\cortext[cor1]{Corresponding author}

\begin{abstract}
\textit{A novel probabilistic approach for the design of mechanical
structures with friction interfaces is proposed. The objective
function is defined as the probability that a specified performance
measure of the forced vibration response is achieved subject to
parameter uncertainties. The practicability of the approach
regarding the extensive amount of required design evaluations is
strictly related to the computational efficiency of the nonlinear
dynamic analysis. Therefore, it is proposed to employ a recently
developed parametric reduced order model (ROM) based on nonlinear
modes of vibration, which can facilitate a decrease of the
computational burden by
several orders of magnitude.\\
The approach was applied to a rotationally periodic assembly of a
bladed disk with underplatform friction dampers. The robustness of
the optimum damper design was significantly improved compared to the
deterministic approach, taking into account uncertainties in the
friction coefficient, the excitation level and the linear damping.
Moreover, a scale invariance for piecewise linear contact
constraints is proven, which can be very useful for the reduction of
the numerical effort for the analysis of such systems.}
\end{abstract}


\end{frontmatter}

\section*{Nomenclature}
\begin{eqnarray}
\nonumber EO && \text{Engine order}\\
\nonumber f && \text{Probability density function}\\
\nonumber \mm f_{\mathrm e} && \text{Excitation force vector}\\
\nonumber \fnl && \text{Nonlinear force vector}\\
\nonumber H && \text{Heaviside function}\\
\nonumber k_{\mathrm n}, k_{\mathrm t} && \text{Normal and tangential contact stiffness}\\
\nonumber m,\moptc && \text{Friction damper mass, optimum mass without uncertainties}\\
\nonumber \mm M, \mm C, \mm K && \text{Mass, damping, stiffness matrix}\\
\nonumber N_{\mathrm h} && \text{Harmonic order of Fourier ansatz}\\
\nonumber p_0, p_{\mathrm n}, \mm p_{\mathrm t} && \text{Initial normal pressure, normal and tangential contact pressure}\\
\nonumber \prob && \text{Probability}\\
\nonumber q && \text{Modal amplitude defined as kinetic energy}\\
\nonumber t && \text{Time}\\
\nonumber \mm u && \text{Displacement vector}\\
\nonumber X,X_{\mathrm{DV}},X_{\mathrm{UC}} && \text{System parameter, design variable, uncertain parameter}\\
\nonumber \stim && \text{Excitation level}\\
\nonumber \eta && \text{Hysteretic damping ratio}\\
\nonumber \lambda, \ommod, \dmod && \text{Eigenvalue, eigenfrequency, damping ratio}\\
\nonumber \mu && \text{Friction coefficient}\\
\nonumber \Omega, \Omega_{\mathrm r} && \text{Excitation frequency, rotational speed}
\end{eqnarray}
\begin{eqnarray}
\nonumber \Theta && \text{Performance measure}\\
\nonumber \dot{\left(~\right)}, \ddot{\left(~\right)} && \text{First and second-order time derivative}\\
\nonumber {\left(~\right)^{\mathrm T}} && \text{Transpose}\\
\nonumber\text{DOF} && \text{Degree of freedom}\\
\nonumber\text{FE} && \text{Finite Element}\\
\nonumber\text{FRF} && \text{Frequency response function}\\
\nonumber\text{HBM} && \text{(High-order) Harmonic Balance Method}\\
\nonumber\text{NMA} && \text{Nonlinear Modal Analysis}\\
\nonumber\text{ODE} && \text{Ordinary Differential Equation}\\
\nonumber\text{ROM} && \text{Reduced Order Model}
\end{eqnarray}

\section{Introduction\label{sec:introduction}}
\paragraph{Friction damping}
Friction damping represents an established technology for the
passive reduction of forced vibrations. Various industrial
applications exist in the field of turbomachinery, where friction
joints are introduced in order to improve the dynamical behavior and
increase the mechanical integrity of bladed disks
\zo{bert1998,pann2003a,petr2006c,char2006a,tatz2013}. Typically, the
goal is to decrease the risk of fatigue by either providing enough
damping in critical operating conditions or shifting
eigenfrequencies away from possible coincidences with the load
spectrum. The design of such friction joints is hampered by the
strongly nonlinear character of the contact constraints, the large
amount of structural degrees of freedom (DOF) required for the
description of the vibration behavior, and the uncertain character
of system parameters.
\paragraph{Robust design under uncertainties}
Uncertainty can be defined as the error between the according
mathematical model and reality \zo{mavr1999}. For friction-damped
mechanical structures, this deviation can occur in material and
geometric properties owing to manufacturing tolerances and material
imperfections. The parameters associated with the description of the
contact dynamics are often difficult to determine and introduce
uncertainty to the system. Similarly, the dynamic effects caused by
the interaction of the underlying mechanical structure with the
surrounding medium can often not be modeled accurately.
Particularly, spatial and spectral distributions of
excitation forces are typical examples for uncertain system parameters.\\
It appears to be common practice in the field of friction damping to
tune the design for only a single set of nominal parameters, \ie to
perform a so-called Single-Point-Optimization, regardless of the
parameter uncertainties \zo{yang1997a,pann2000b,laxa2007a}. If the
variability of uncertain parameters is not properly taken into
account in the design process, even high-fidelity models cannot
yield optimum designs with respect to robustness. Hence, this
localized optimization can result in a design that is highly
sensitive to even slight changes
in the parameters \zo{huyse2002}.\\
Several authors addressed uncertainty in terms of non-probabilistic,
post-optimality sensitivity analysis. Cameron \etal \zo{came1990}
introduced the terminology of the damper performance curve and
employed this strategy to find a design point that is widely
reliable with respect to variations in the excitation level and the
linear damping. Cha and Sinha \zo{cha2003} numerically determined
the optimum normal preload of a friction damper that is largely
insensitive regarding different types of excitation. Several
approaches are known for the analysis of the stochastic
characteristics of the nonlinear forced response for stochastically
distributed parameters \zo{petr2007b,petr2009a,kumar2010}, but they
have not been applied to reliability-based design so far.
\paragraph{Reduced order modeling}
Probabilistic approaches typically involve integrals over the
relevant parameter domain. The evaluation of these integrals cannot
be performed analytically in closed form in most cases and, thus,
has to be performed numerically. This task often involves
evaluations of the design performance for a large number of
parameter points. This can be computationally expensive owing to the
nonlinear character and considering large problem dimensions for the
dynamic analysis. The practicability of probabilistic approaches for
friction-damped structures therefore relies on the efficiency of the
dynamic analysis. It is thus essential to employ reduced order
modeling (ROM) strategies in order to decrease the otherwise
prohibitive computational costs.\\
State-of-the-art strategies to reduce the computational effort for
the dynamic analysis in this field include component mode synthesis
\zo{crai2000}, (multi-)harmonic balance \zo{pier1985} and direct
tracing of resonances \zo{petr2006b}. However, there still remains a
demand for a more drastic order reduction in conjunction with
probabilistic design approaches.\\
The concept of nonlinear modes can generally be employed to extract
the essential signature of oscillatory nonlinear systems
\zo{vaka2008,kers2009} and therefore gives rise to order reduction
opportunities for nonlinear systems \zo{touz2006,blan2013}. In Krack
\etal \zo{krac2013a,krac2013d}, an approach was proposed that
reduces the computational effort by several orders of magnitude
within reasonable ranges of validity. However, it cannot directly be
employed in a sophisticated design process since the ROM lacks of an
appropriate parameter space.
\paragraph{Proposed approach and outline}
The objective of this study is twofold: Firstly, a novel reliability
optimization strategy for the design of friction-damped mechanical
structures under uncertainties is proposed. Secondly, the
application of an extremely efficient parametric ROM based on
nonlinear modes in course of sophisticated probabilistic design
approaches is proposed.\\
This paper shall be regarded as a proof-of-concept study for the
proposed overall design approach. Hence, only a comparatively low
number of uncertain parameters was considered in the numerical
example. Furthermore, the methodology was only applied to systems
driven by purely harmonic forcing in vicinity of a single resonance.
While this is often a reasonable assumption, it is also in
accordance with the current limitations of the employed ROM which
can currently treat neither generic excitation frequency spectra nor
nonlinear modal interactions. An important goal of this article is
to apply the design approach to an example of realistic complexity
in terms of structural and contact dynamics of the intended
industrial applications.\\
This paper is organized as follows: The description of the
structural dynamic behavior is briefly addressed in \sref{methods}
with an emphasis on the formulation of the contact constraints and
the novel scale invariance exploited for the parameter studies. The
reliability optimization approach is detailed in
\sref{robust_design}. The nonlinear modal reduced order model is
presented in \sref{nmrom} with a focus on the parameter space
extended in this work. An overview of the overall proposed approach
is provided in \sref{overview}. Finally, the methodology is applied
to a bladed disk with underplatform friction dampers in
\sref{numexample}.

\section{Modeling\label{sec:methods}}
\subsection{Dynamic description of structures with contact interfaces\label{sec:structdyn}}
The dynamics of a spatially discrete, mechanical system is governed
by a system of second-order ordinary differential equations,
\e{\mm M \mm{\ddot u}(t) + \mm C\mm{\dot u}(t) + \mm K\mm u(t) +
\fnl\left(\mm u(t),\mm{\dot u}(t)\right) = \mm f_{\mathrm
e}(t)\fp}{eqm}
Herein, \g{\mm M=\mm M\tra} is the symmetric positive definite mass
matrix. \g{\mm C} and \g{\mm K} are the damping and the stiffness
matrix, respectively. \g{\fnl\left(\mm u(t),\mm{\dot u}(t)\right)},
and \g{\mm f_{\mathrm e}(t)} are vectors of nonlinear and excitation
forces. \g{\mm u} is the vector of generalized coordinates.\\
Temperature and wear effects may generally alter the material
properties and the geometry of friction-damped systems. These
effects typically occur on timescales that are much longer than the
one associated with the vibration behavior. It is thus reasonable to
consider the structural matrices and nonlinear forces in \eref{eqm}
as time-invariant. As generalized coordinates are allowed in
\eref{eqm}, it is possible to apply component mode synthesis in
order to reduce the dynamics of the system to the most relevant
linear modes. This can be useful when dealing with large-scale
structures with only localized contact joints. In this study, only
the nonlinear forces associated with the contact constraints are
treated as nonlinear, while all other possible sources of
nonlinearity such as large deformations or nonlinear
fluid-structure-interaction are not taken into account in the
dynamic analysis.

\subsection{Contact modeling\label{sec:contact}}
Unilateral contact in normal direction and dry Coulomb friction in
the tangential plane are considered in this study. In the field of
friction damping, it is common to apply elastic formulations of
these contact laws and to interpret the associated elastic
properties physically as contact stiffness
\zo{john1989,siew2006a,petr2003b,firr2011,krac2013b}.
Mathematically, the resulting constitutive laws are equivalent to
penalty re-formulations of the complementarity conditions with
constant normal and tangential stiffness parameters \g{k_{\mathrm
n}} and \g{k_{\mathrm t}} \zo{wrig2006},
\ea{p_{\mathrm n}(t) &=& \left(k_{\mathrm n} u_{\mathrm n}(t)+
p_0\right)_+\,\label{eq:pnormal}\\
\mm p_{\mathrm t}(t)&=&\begin{cases}\mm 0 & \text{lift-off}\\
k_{\mathrm t}\left(\mm u_{\mathrm t}(t)-\mm u_{\mathrm
t}\left(t_{\mathrm{stick}}\right)\right)+\mm p_{\mathrm
t}\left(t_{\mathrm{stick}}\right) & \text{stick}\\
\mu p_{\mathrm n}(t)\frac{\mm{\dot u}_{\mathrm t}(t)}{\|\mm{\dot
u}_{\mathrm t}(t)\|} & \text{slip}\end{cases}\fp}{nttc}
Herein, \g{t_{\mathrm{stick}}} is the time instant at which the
current stick phase began and \g{\mm u_{\mathrm
t}\left(t_{\mathrm{stick}}\right), \mm p_{\mathrm
t}\left(t_{\mathrm{stick}}\right)} are the values just before the
current stick phase began. The contact model is formulated in terms
of the contact surface normal and tangential pressures \g{p_{\mathrm
n}}, \g{\mm p_{\mathrm t}}, respectively. The friction coefficient
\g{\mu} is assumed to be constant and identical for stick and slip
state. The initial normal pressure \g{p_0} can be negative to
account for initial clearances, see \eg \zo{petr2003b} for details.
The contact model in \erefs{pnormal}{nttc} separately models stick,
slip and lift-off states and accounts for the effect of the normal
dynamics on the friction behavior. The contact pressures have been
integrated over the contact interfaces in order to obtain the vector
of nonlinear forces \g{\fnl} in \eref{eqm}. Therefore, a
node-to-node formulation was used with the matching nodes of the
surface elements as integration points. Linear surface elements have
been used in the case study, such that the surface integral was
approximated by a weighted sum over the integration points, where
the weights represent the area associated with each integration
point \zo{wrig2006}. Small relative deflections were assumed for
this procedure such that nonlinearity stems solely from the contact
laws, but not from the kinematics or the pressure integration scheme.\\
A scale invariance is exploited in this study to determine the
amplitude-dependent nonlinear modes for different scaling values
\g{\scal} of the initial pressure distribution \g{\scal p_0} from
the nonlinear modes computed for a single value \g{p_0}. This scale
invariance can be stated as follows:
\ea{\nonumber\text{\underline{scale invariance}:}&\\
\nonumber \text{If } \left[\mm u\uudist\mm{\dot u}\right](t) &\text{ is a
solution to \eref{eqm} for }
\lbrace\left[\mm u\uudist\mm{\dot u}\right]_{(t=0)},p_0,\mm f_{\mathrm e}(t)\rbrace\fk\\
\text{ then } \scal\left[\mm u\uudist\mm{\dot u}\right](t) &\text{
is a solution to \eref{eqm} for } \lbrace\scal\left[\mm
u\uudist\mm{\dot u}\right]_{(t=0)},\scal p_0,\scal\mm f_{\mathrm
e}(t)\rbrace\fp}{scaling_property}
Starting from a specific solution \g{\left[\mm u\uudist\mm{\dot
u}\right](t)} corresponding to initial value \g{\left[\mm
u\uudist\mm{\dot u}\right]_{(t=0)}}, preload \g{p_0} and loading
\g{\mm f_{\mathrm e}(t)}, this scale invariance allows to determine
the solution to a scaled problem without the need for
re-computation. A similar scale invariance was reported for the
special case of constant normal loading in \zo{sext2002b}. It was
postulated in \zo{krac2013a} for the more general case of variable
normal contact force including lift-off. A formal proof can be found
in \arefo{proof_scaling} valid
for the contact formulation applied in this paper.\\
The exploitation of the scale invariance can significantly increase
the computational efficiency of the design process. This feature
also provides a better understanding of the often reported analogy
between the increase of the excitation and the decrease of the
normal preload \zo{pann2002b}. It is therefore believed that this
scale invariance can be applied for a better understanding and
computational effort reduction in nonlinear dynamic analyses of
friction-damped structures far beyond the application to
nonlinear modes.\\
It should be noticed that the preload scaling \g{\scal} can not
always be considered as a directly accessible parameter. In case of
friction-damped rotating bladed disks, the preload is typically
influenced by the pretwist design in case of shroud interfaces
\zo{szwe2003} and adjustment of damper density and geometry in case
of underplatform dampers \zo{pann2003a}. The actual preload also
depends on the rotational speed and the mass and stiffness
distributions of the resulting assembly.\\
Finally, it should be emphasized that contact laws such as piecewise
nonlinear \zo{pann2002b} or Lagrangian \zo{naci2003} formulations
could be easily applied to the proposed methodology. However, the
scale invariance will then no longer be valid in general. Hence, the
dimension of the parameter domain for which the dynamic analyses
have actually to be carried out, would increase by one.

\section{Robust design optimization\label{sec:robust_design}}
The appropriate formulation of the optimization problem for a system
with uncertainties is not a trivial task. Considering the
performance sensitivity in addition to the performance \g{\perf} at
the design point generally yields a multi-objective optimization
problem. The solution of such problems is typically much more
computationally expensive compared to single-objective problems and
leads to a whole set of Pareto-optimal designs rather than a single
optimum design. Instead of formulating a multi-objective problem,
the robustness can also be incorporated into a single-objective
criterion employing the Von-Neumann-Morgenstern statistical decision
theory \zo{huyse2002}. The optimization problem is then defined as
the maximization of the expected value of the reliability, \ie the
`probability of success' \g{\prob\left[\perf\left(\mm X\right)\geq
\perfref\right]} that a specified performance \g{\perfref} is
achieved \zo{huyse2002,krac2012a,krac2013c},
\ea{\nonumber \text{maximize} & \prob\left[\perf\left(\mm
X\right)\geq
\perfref\right]\\
\nonumber \text{with respect to} & \mm X_{\mathrm{DV}}\\
\text{subject to uncertainties in} & \mm
X_{\mathrm{UC}}\fp}{design_problem}
The entire set of system parameters \g{\mm X} comprises the set of
design variables \g{\mm X_{\mathrm{DV}}} as well as the uncertain
parameters \g{\mm X_{\mathrm{UC}}}. Design variables may also be
treated as uncertain. Thus, the system parameters are the union of
design and uncertain parameters,
\e{ \mm X = \mm X_{\mathrm{UC}}\cup\mm
X_{\mathrm{DV}}\fp}{system_parameters}
The probability of success in \eref{design_problem} can be expressed
as expectation integral over the entire domain of all
\g{N_{\mathrm{UC}}} uncertain parameters,
\ea{\nonumber\prob\left[\perf\left(\mm X\right)\geq
\perfref\right]=\quad\quad\\
\intl{X_{N_{\mathrm{UC}},\min}}{X_{N_{\mathrm{UC}},\max}} \cdots
\intl{X_{1,\min}}{X_{1,\max}} H\left[\perf\left(\mm X\right) -
\perfref\right]f_1(X_1)\dd X_1\cdots
f_{N_{\mathrm{UC}}}(X_{N_{\mathrm{UC}}})\dd
X_{N_{\mathrm{UC}}}\fp}{probability_of_success}
Herein, \g{H} is the heaviside function, \g{f_i} are the probability
density functions of the uncertain parameters \g{X_i\in\mm
X_{\mathrm{UC}}}.
\fss[tbh]{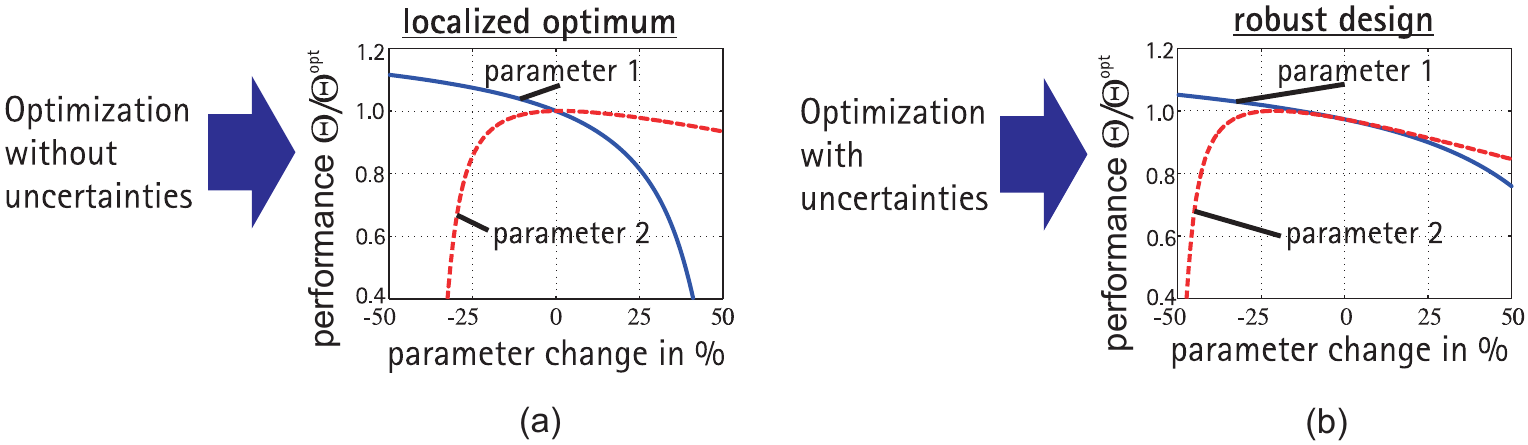}{Qualitative comparison between localized
optimization and robust design (~(a) localized optimization, (b)
robust design~)}{1.0}
\\
Owing to the integral formulation of \eref{probability_of_success},
this approach explicitly accounts for off-design performance. The
optimum design will therefore deviate from the localized optimum
obtained by simply maximizing performance for nominal parameters.
Typically, the performance of the robust design is slightly worse
for nominal parameters but exhibits a greater robustness against
considerable parameter changes, see \fref{figure01}.
\eref{probability_of_success} represents the limit case of a
weighted-sum reformulation of a multi-objective problem with the
qualities at every design point as the objectives and the
corresponding probability values as weighting factors.\\
An appropriate performance measure \g{\perf} needs to be selected
for the design process. For the sought application to friction
damping, conceivable measures are displacement amplitudes, dynamic
stresses, strength ratios, resonance frequency or combinations of
these measures. The performance threshold \g{\perfref} could be
derived from strength or fatigue analyses.\\
The evaluation of the probability integral can be performed
analytically in some cases. In particular, if it is valid to
approximate the performance \g{\perf(\mm X)} in terms of a
linearized function in the parameters \g{\mm X}, analytical
expressions for the reliability can be easily obtained
\zo{krac2012a}. In general, however, the integration needs to be
carried out using numerical integration methods. The adaptive
recursive Simpson's integration rule was employed in this study
\zo{davis1967}. This is a conventional integration method available
in most engineering software frameworks. For this method, the
computational cost increases exponentially with \g{N_{\mathrm{UC}}}.
In this paper, at most three uncertain variables had to be taken
into account simultaneously, which still led to acceptable
computational effort. When more uncertain parameters are taken into
account, this integration method can soon become prohibitive. In
this case, Monte-Carlo or sparse grid methods typically perform
better \zo{pfaf2012}. However, those methods are more complex and
feature additional parameters that need to be specified with care.
Hence, those methods have not been further considered within this
paper.

\section{Nonlinear modal reduced order model\label{sec:nmrom}}
The reliability optimization proposed in this paper could be
prohibitive even for the most efficient quadrature scheme to
evaluate \eref{probability_of_success}, if the cost for the
performance evaluation \g{\perf(\mm X)} for a single parameter point
is too large. In order to avoid this, the ROM developed in
\zo{krac2013a} is used in this paper, which is based on the concept
of nonlinear modes. The benefit of this ROM is that it is extremely
cheap to evaluate since it leads to a scalar nonlinear algebraic
equation for each parameter point. The downside is that the validity
is restricted to those dynamic regimes, where the
vibration energy is mainly confined to a single nonlinear mode.\\
In order to employ the ROM in the reliability design process, the
generally parameter-dependent nonlinear modes have to be available
in the whole parameter domain \g{\mm X_{\min}\leq\mm X\leq\mm
X_{\max}}, see \eref{probability_of_success}. For a group of
parameters, a costly re-computation of the nonlinear modes can be
avoided. This applies in an exact manner to the normal preload due
to the scale invariance, see \ssref{contact}. This also applies
approximately to parameters associated to near-resonant harmonic
forcing and weak linear damping. The re-computation of the nonlinear
modes is avoided by assuming that the influence of these terms on
the geometry of the invariant manifold can be neglected and the
modal properties can be approximated by the nonlinear modes computed
for a surrogate system without these terms \zo{krac2013a}.\\
It should be stressed that preload, damping and forcing parameters
are often regarded as uncertain and/or relevant design variables in
the field of friction damping \zo{laxa2008a,petr2009a,krac2013d}.
Hence, the proposed parameter treatment strategy can greatly enhance
the computational benefit of the nonlinear modal ROM in the course
of reliability optimization of friction-damped structures. For the
remaining group of parameters such as contact parameters, material
properties, geometric properties or temperature, the nonlinear modes
generally have to be re-computed.

\section{Overview of the proposed reliability analysis
\label{sec:overview}}
\fss[tbh]{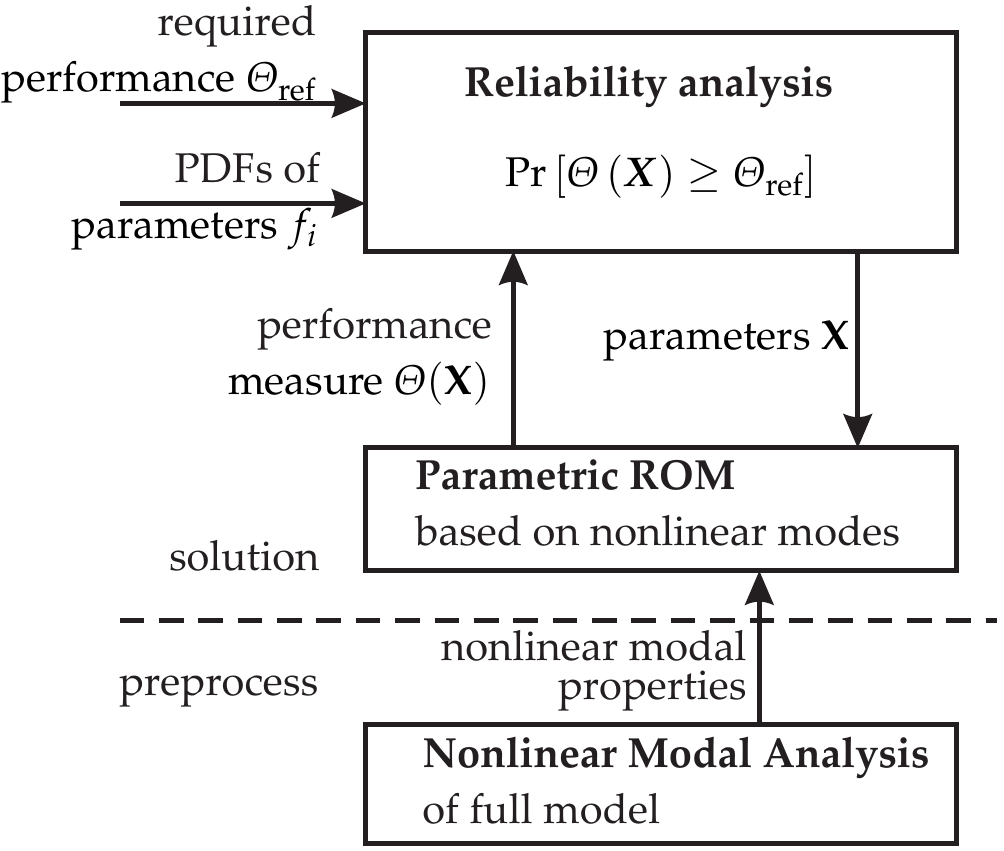}{Flow chart of the reliability analysis based on
the nonlinear modal ROM}{0.7}
A flow chart of the resulting approach is illustrated in
\fref{figure02}. It consists of two parts, namely the Nonlinear
Modal Analysis (NMA as preprocessing part and the
solution part that addresses the actual reliability analysis.\\
The NMA is first carried out as described in \sref{nmrom} in order
to compute the modal properties, which are required for the ROM
creation. These modal properties depend may depend on several system
parameters later have to be varied in course of design optimization.
The modal properties have to be computed with respect to only those
parameters that are not inherently included in the ROM. For the
numerical example in \sref{numexample} this was the case for the
friction coefficient.\\
In the solution step, the constructed ROM can then be utilized for
the efficient prediction of the vibration behavior. This prediction
has to be allowed for arbitrary points \g{\mm X} in the parameter
space. These points in general do not coincide with the ones for
which the modal properties have been computed in the preprocessing
step. To this end, a piecewise cubic interpolation between the
points available from the NMA was carried out as described in
\zo{krac2013a}.\\
The reliability can then be computed by numerical integration of the
integral in \eref{probability_of_success}. The integration scheme
selects the required sampling points \g{\mm X} based on the
specified integral limits and the performance measures returned by
the ROM (black box). The reliability analysis can further be
embedded within an optimization procedure in order to actually solve
the design problem stated in \eref{design_problem}.

\section{Numerical example\label{sec:numexample}}
\subsection{Problem definition\label{sec:problem}}
\fss[th]{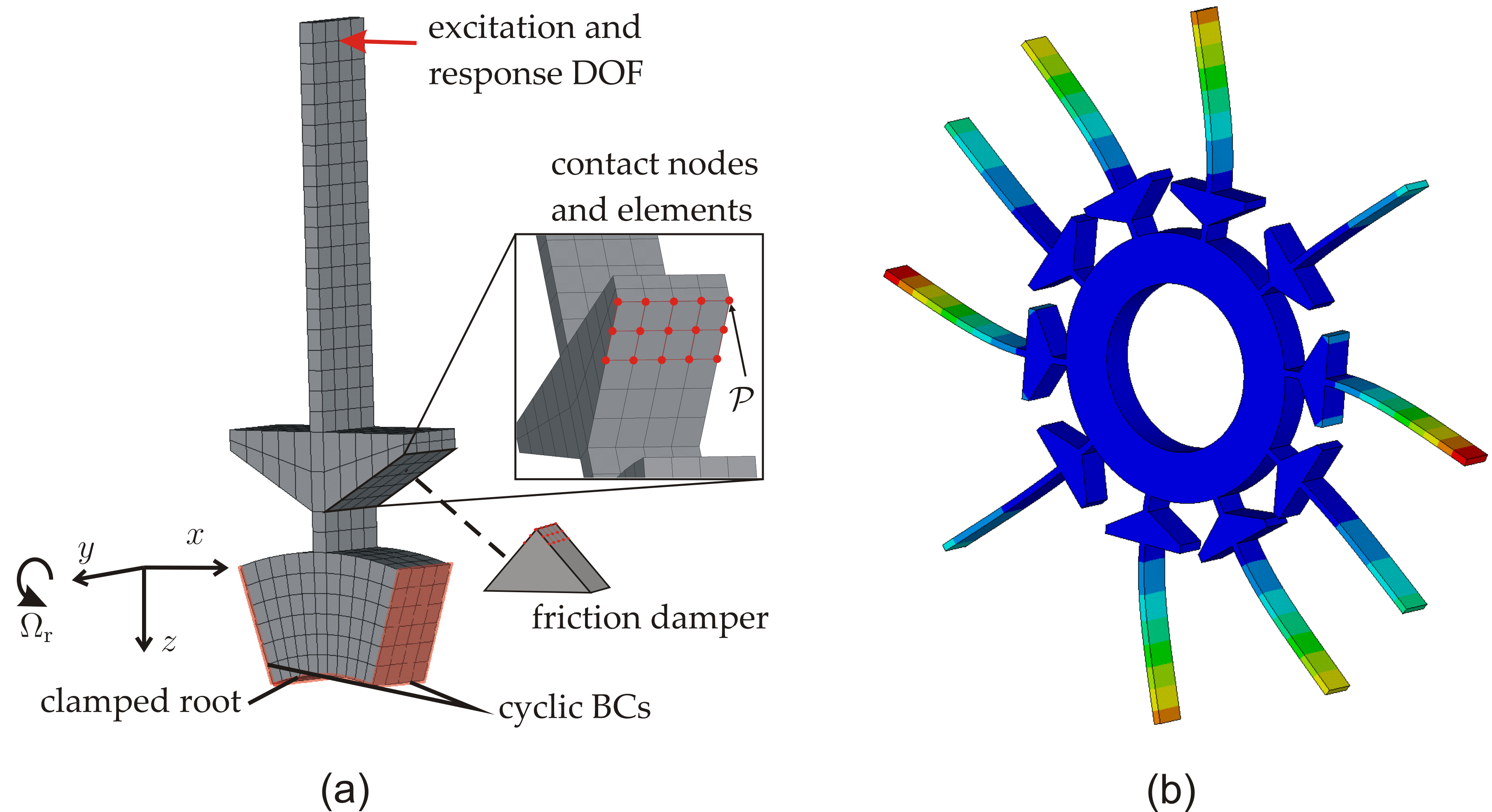}{Finite element model of a rotating bladed disk
with underplatform dampers (~(a) cyclic segment with boundary
conditions, (b) investigated mode shape without friction
dampers~)}{1.0}
The proposed reliability optimization strategy was applied to the
underplatform damper design of the rotating bladed disk depicted in
\fref{figure03}a. In this proof-of-concept study, several
simplifications were made for the model.
\begin{itemize}
\item \textit{Cyclic symmetry}: The bladed disk was considered as
perfectly tuned. This is a strong restriction of the uncertain
parameter domain since this excludes mistuning, \ie blade-by-blade
variations of parameters. Mistuning can in general have an essential
effect on energy localization and overall vibration behavior. It
should be emphasized that mistuning is mainly relevant for the
special application of rotationally periodic structures. Moreover,
this simplification facilitates a more detailed analysis of the
individual qualitative influences of other parameters.
\item \textit{Harmonic excitation}: A near-resonant traveling-wave
type forcing of engine-order $EO=2$ was imposed. The excitation
frequency \g{\Omega} is thus directly related to the rotational
speed \g{\Omega_{\mathrm r}} by \g{\Omega=EO\Omega_{\mathrm r}}. It
was assumed that the resonant response exhibits the particular
symmetry in accordance with the symmetry of excitation and structure
\zo{siewert2008}. This is not necessarily the case for nonlinear
cyclic assemblies \zo{king1995,geor2008}.
\item \textit{Academic geometry}: The geometry in
\fref{figure03}a is a caricature of a realistic geometry of a
turbomachinery bladed disk.
\item \textit{Simplified damper modeling}: The underplatform wedge
dampers were assumed to be rigid, which is a common simplification.
The damper mass has a significant influence on the normal preload in
the contact interfaces, which in turn has an essential effect on the
damping performance. This influence on the normal preload was taken
into account in the present case study. In practice, the variation
of the damper mass is achieved by variation of damper geometry or
material properties. In general these modifications have an
influence on inertial and elastic forces within the assembly, and
may affect its dynamic behavior. These influences were found to play
a minor role for the relevant mass ranges in the considered
parameter studies. This has also been found by other researchers
\zo{sanl2001,jare2001}. Hence, only the nominal damper with mass
\g{\moptc}, as defined later, was considered for the derivation of the
structural dynamical properties. During variation of the mass in the
following parameter study, only its influence on the normal preload
was therefore taken into account.
\item \textit{Simplified centrifugal effects}:
The normal preload of the contact interfaces was determined from the
static equilibrium with the centrifugal force
\g{mr_{\mathrm{cg}}\Omega_{\mathrm r}^2}, where \g{r_{\mathrm{cg}}}
is the distance from the rotation axis to the center of gravity of
the dampers. Moreover, the structural matrices of the assembly were
regarded as constant within the considered rotational speed range.
Hence, centrifugal stiffening was only taken into account for
nominal rotor speed of the bladed disk without dampers.
\item \textit{Discrete forcing and proportional damping}:
In state-of-the-art forced response analyses, the fluid-structure
interaction is commonly modeled in terms of external forcing and
aerodynamic damping. The external forces are caused by the rotation
of the bladed disk in the inhomogeneous (steady) fluid pressure
field. Instead of applying an actual fluctuating pressure field, a
discrete excitation force at the location indicated in
\fref{figure03}a was specified in this case study. The aerodynamic
damping was approximated as weak proportional damping \g{\mm
C=\ii\eta\mm K}.
\item \textit{Fixed boundary conditions}: The disk was considered
to be fixed to a rigid hub without consideration of rotor unbalance
or multistage influence. Gyroscopic influences were neglected.
\end{itemize}
The first two simplifications could only be released by extending
the proposed ROM approach to non-harmonic forcing and nonlinear
modal interactions. The remaining simplifications are assumed not to
have a significant qualitative influence on the vibration behavior
and could be easily taken into account for an actual industrial
case study.\\
Structural and underplatform damper contact dynamics were modeled
employing a FE model and 3D contact constraints imposed at discrete
contact interfaces, \cf \fref{figure03}a. This is crucial in order
to obtain a complexity of the nonlinear dynamic analysis that is
realistic for industrial applications
\zo{krac2012b,siew2009a,zucc2013}. A homogeneous distribution of the
initial normal pressure at the contact interface was assumed. It is
conjectured that a more comprehensive modeling of the initial normal
pressure distribution would not affect the dynamic analysis
complexity, which was the primary motivation for the proposed model
reduction approach.\\
For the design problem, only the first mode of spatial harmonic
index \g{2} was investigated. The deformed mode shape of this mode
is illustrated in \fref{figure03}b. The performance measure was
specified as the maximum vibration amplitude, \g{a}, of the response
DOF indicated in \fref{figure03}a. The damper mass \g{m} was
considered as the only design variable. Uncertainties in the
excitation level \g{\stim}, the linear damping \g{\eta} and the
friction coefficient \g{\mu} were taken into account. The design
problem can thus be stated as
\ea{\nonumber \text{maximize} &
\prob\left[a\left(\mm X\right)\leq\areference\right]\\
\nonumber \text{with respect to} & X_{\mathrm{DV}}=m\\
\text{subject to uncertainties} & \mm
X_{\mathrm{UC}}=\left[\stim\,\,\eta\,\,\mu\right]^{\mathrm
T}\fp}{design_problem_specific}
The nominal parameter values \g{\mm X_0 =
\left[m_0\,\stim_0\,\eta_0\,\mu_0\right]\tra} were specified as
\g{\mm X_0 = \matrix{cccc}{\moptc & 1 & 0.15\% & 0.3}\tra}, where
\g{\moptc} refers to the mass that minimizes the response \g{a} for
otherwise nominal parameters.\\
Arbitrary probability density functions (PDFs) \g{f_i} are allowed
in \eref{probability_of_success}. In practice it is certainly not
trivial to determine realistic PDFs \g{f_i} by means of
measurements, error estimations and experience. Hence, the
parameters that describe the PDFs will often be uncertain
themselves. It is thus interesting to vary these parameters and
estimate their influence on the robust design \zo{krac2012a}. In
this study, a gaussian stochastic distribution was assumed for all
parameters. Hence the probability integral in
\eref{probability_of_success} would have to be evaluated on an
infinite domain. In this study, the numerical integration is
restricted to the \g{3-\sigma-}range corresponding to the
\g{99.73}rd percentile of the parameter domain. It is important to
note that the gaussian distribution is of course also defined for
negative parameters, which might lead to questionable results in
case of \eg the friction coefficient. The variabilities investigated
in this study, however, lead to negligible probability for negative
parameter values in general.

\subsection{Nonlinear modal properties\label{sec:nmprop}}
\fss[tbh]{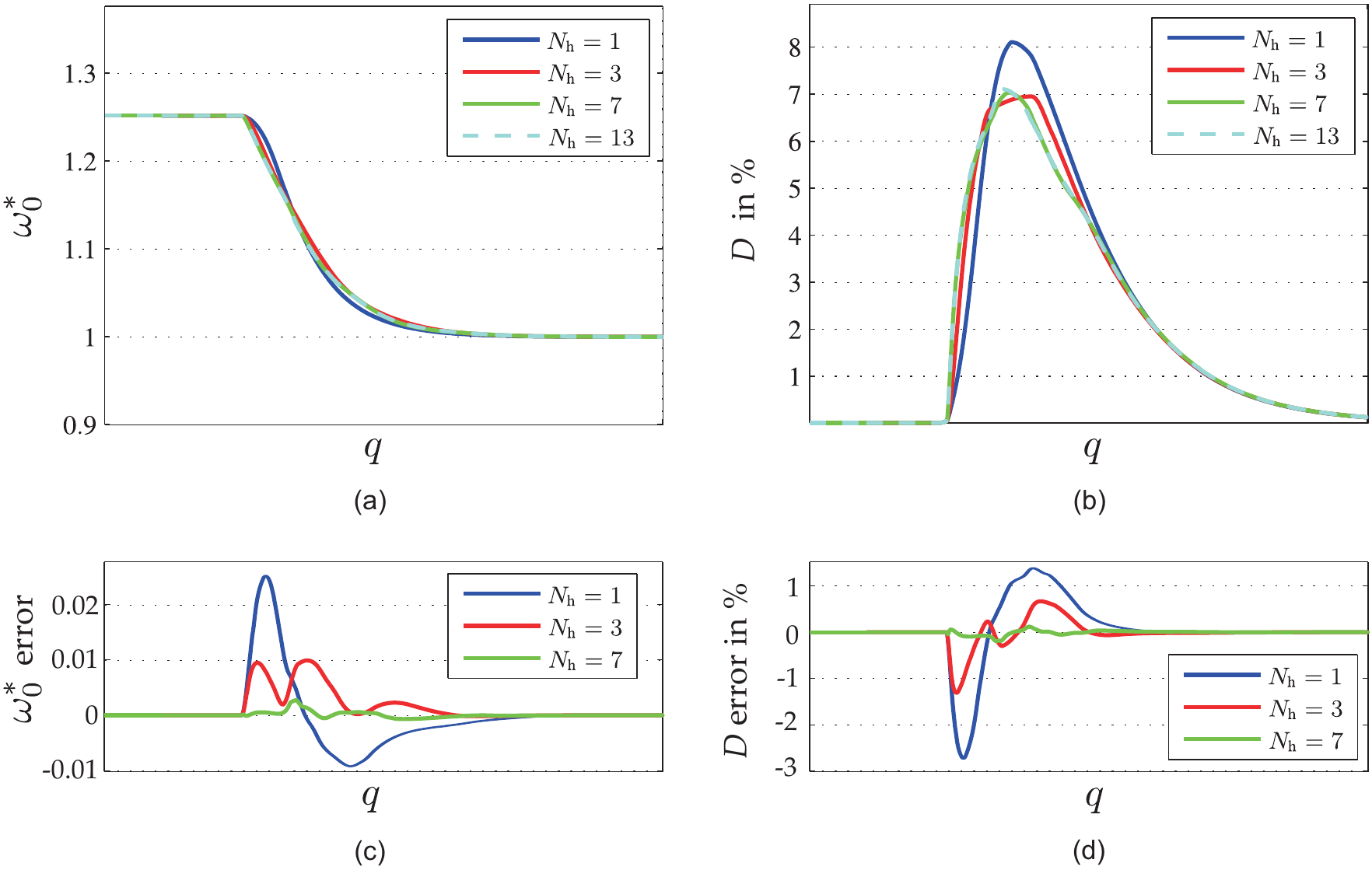}{Convergence of modal properties with respect to
harmonic order \g{\nh} (~(a) eigenfrequency, (b) modal damping
ratio, (c) eigenfrequency error compared to \g{\nh=13}, (d) modal
damping ratio error compared to \g{\nh=13}~)}{1.0}
\paragraph{Results for nominal parameters}
The nonlinear modal properties for the considered mode have at first
been computed for nominal parameters and are illustrated in
\fref{figure04}. Throughout this section, variables indicated by an
asterisk (\g{{}^*}) were normalized to the forced resonant case
without dampers for nominal parameters. The modal amplitude \g{q}
refers to maximum kinetic energy. The typical friction damping
effect can be ascertained from the variation of the modal properties
with the modal amplitude: When the entire contact area is stuck,
there is no friction damping and the eigenfrequency is constant. For
moderate amplitudes, a maximum damping value is reached in the
microslip regime and a softening effect can be deduced from the
decreasing eigenfrequency. For large amplitudes, the modal damping
decreases again and the system
asymptotically approaches quasi-linear behavior.\\
Several harmonics are required to reach asymptotic convergence of
the modal properties. A harmonic order \g{\nh=7} was considered as
sufficient regarding the modal analysis results and the small error
compared to \g{\nh=13}, see \frefs{figure04}c-d. Consequently all
integer harmonics from the zeroth to the seventh were retained for
all further investigations, including the HBM reference
computations. In addition to the zeroth and fundamental harmonic,
all harmonics up to order were at least locally in the vicinity of
the contact interfaces, significant contributions of the higher
harmonics are induced by the nonlinear contact forces. This can also
be seen from the invariant manifold depicted in \fref{figure05} in
the coordinates of the contact node \g{\mathcal P} (see
\fref{figure03}a). It is generally interesting to investigate the
spatial distribution of the contact state. Owing to the simplicity
of the geometry and the assumed homogeneous normal pressure
distribution, however, the contact dynamics and states were found to
be largely homogeneous. Hence, the spatial contact distribution was
not illustrated in this paper.
\fss[th]{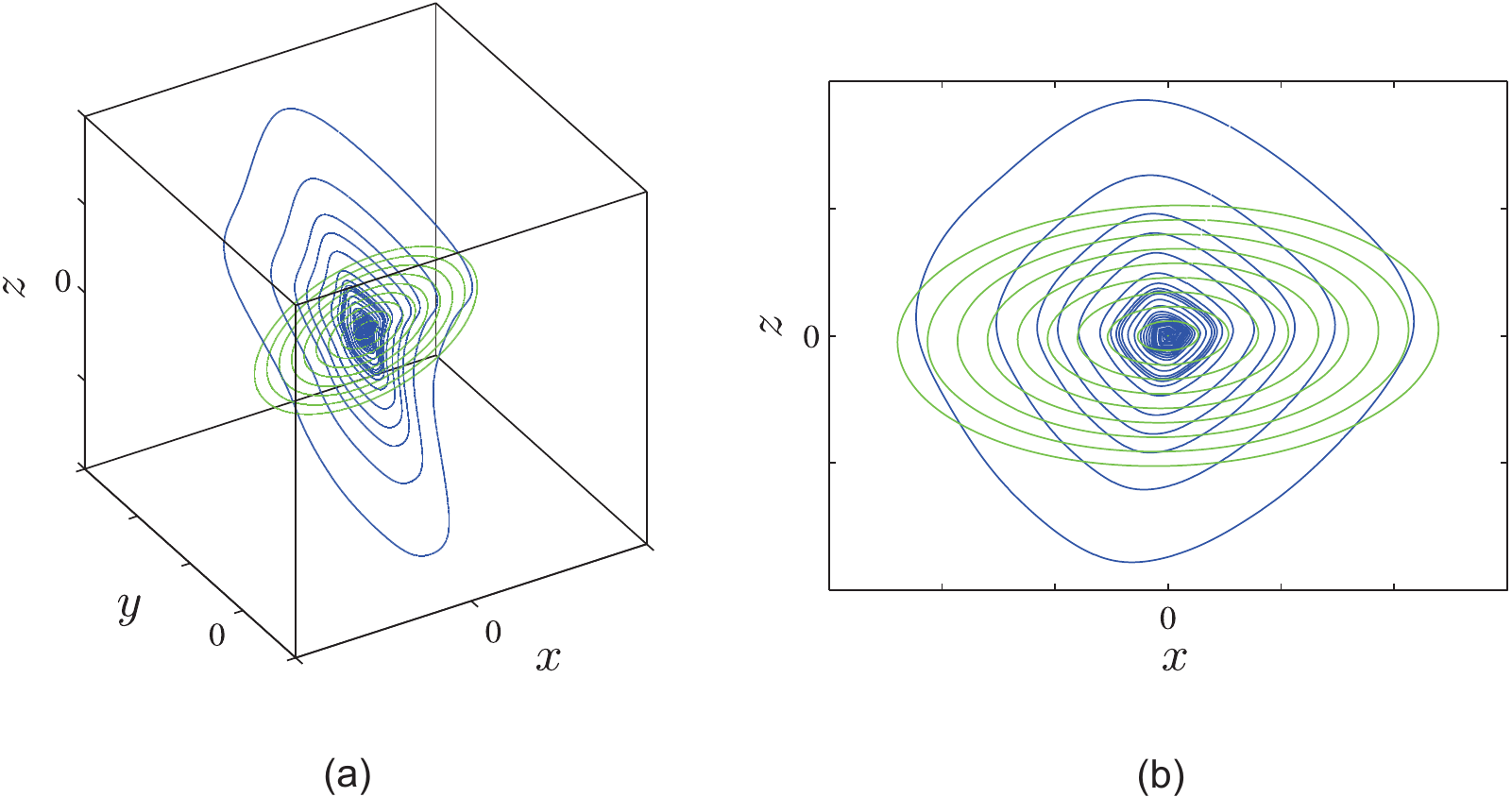}{Invariant manifold of nonlinear mode represented
as orbits of contact node $\mathcal P$; lightly plotted curves
corresponds to the linear mode for entirely stuck contact conditions
(~(a) three dimensional representation, (b) two dimensional
representation~)}{1.0}
\paragraph{Influence of the friction coefficient}
\fss[th]{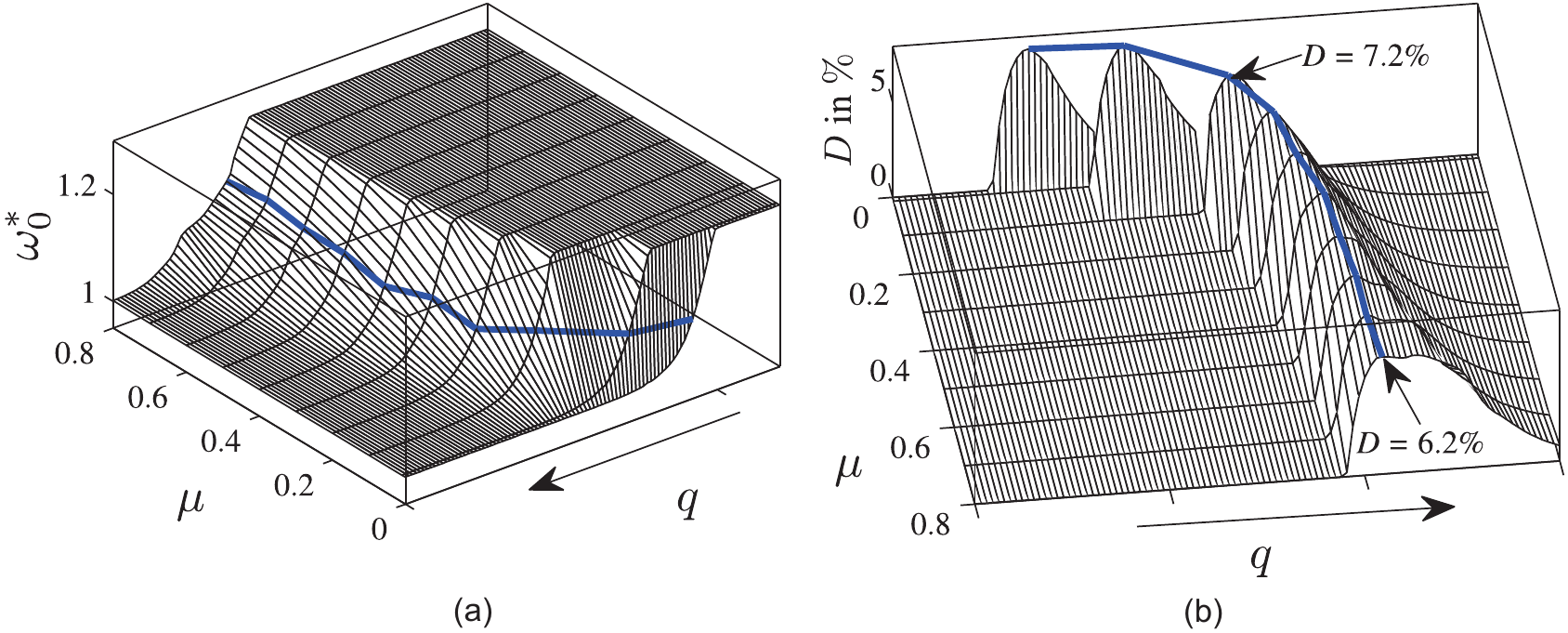}{Modal properties as a function of modal amplitude
and friction coefficient (~(a) eigenfrequency, (b) modal damping
ratio~)}{1.0}
In order to conduct the reliability optimization, the nonlinear
modes not only have to be available for nominal parameters but for
the entire parameter domain spanned by the system parameters \g{\mm
X = \left[m\,\stim\,\eta\,\mu\right]\tra}. Owing to the assumptions
described in \ssref{problem}, the mass \g{m} was varied without
re-computation of the modal basis by exploiting the scale invariance
as derived in \ssref{contact}. Also, the forcing amplitude \g{\stim}
and the weak linear damping \g{\eta} are inherently included in the
parameter space of the ROM. Only for the friction coefficient
\g{\mu}, the nonlinear modes had to be re-computed.\\
The modal properties are depicted with respect to the friction
coefficient and the modal amplitude in \fref{figure06}. For a
friction contact with constant normal load, the characteristic would
only be shifted along the modal amplitude axis. In this case
however, the normal dynamics influence the stick-slip transitions.
For larger friction coefficients, also larger relative motion
amplitudes are required to achieve the same amount of slipping. For
larger amplitudes, however, also longer lift-off phases occur during
one vibration cycle in case of variable normal loading. Thus, the
maximum damping capacity is decreased for larger friction
coefficients.

\subsection{Forced response and sensitivities\label{sec:frf}}
\fss[th]{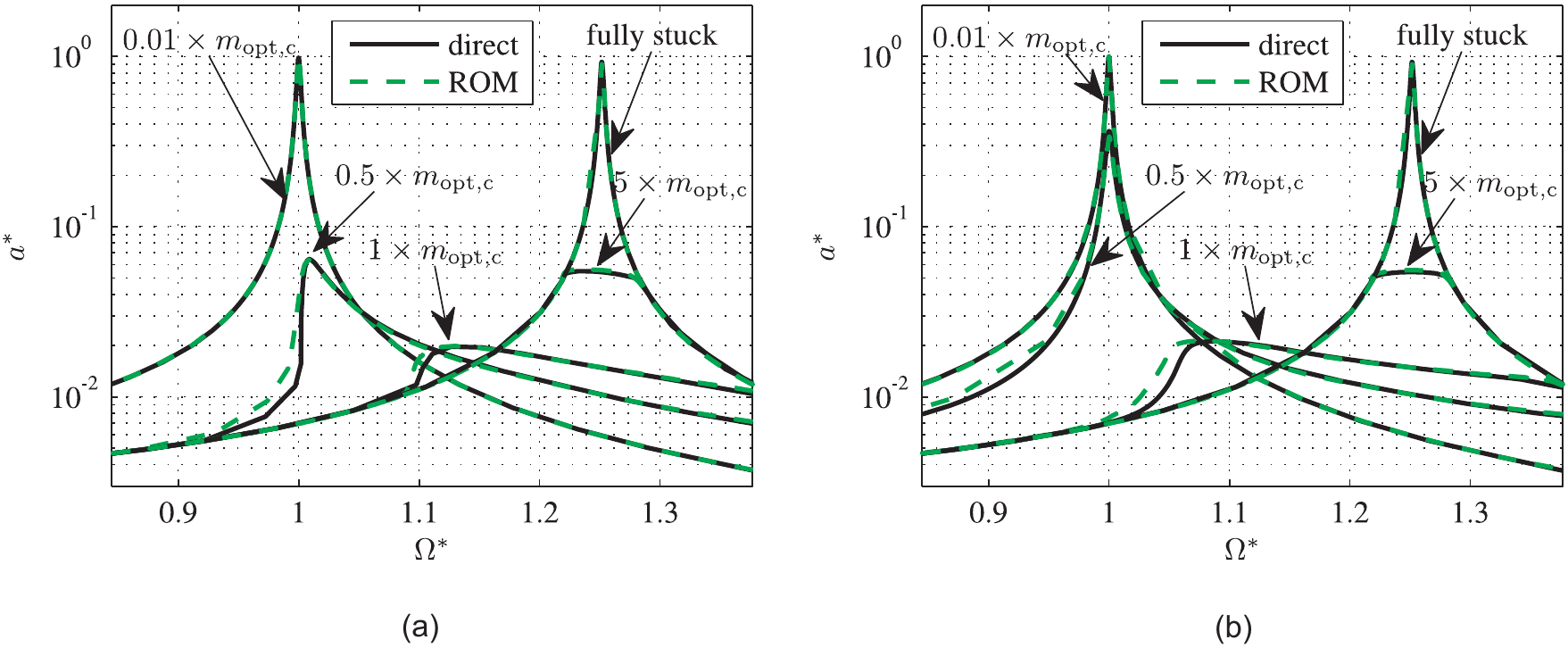}{Frequency response functions for different damper
mass values (~(a) normal preload treated as fixed, (b) normal
preload increasing quadratically with rotational speed~)}{1.0}
In order to assess the accuracy of the ROM compared to direct HBM,
the frequency response functions are depicted in \fref{figure07} for
different damper mass values. For \fref{figure07}a, the normal
preload has been fixed to the value corresponding to the centrifugal
force at nominal rotor speed \g{\Omega_{\mathrm r}^*}. For
\fref{figure07}b, the influence of the rotor speed
\g{\Omega_{\mathrm r}\neq\Omega_{\mathrm r}^*} on the normal preload
was taken into account. Apparently, the effect of the
speed-dependent preload is significant in this case. Hence, this
effect was considered for the analyses presented in the sequel of
this paper. Generally, a very good agreement can be ascertained
between direct and ROM computations. This applies in particular to
the results in vicinity of resonance which are of primary interest
for the reliability. The resonant amplitudes according to the ROM
were found to deviate from the direct HBM reference by less than
\g{1\%} for the results depicted in \fref{figure07}.
\fss[th]{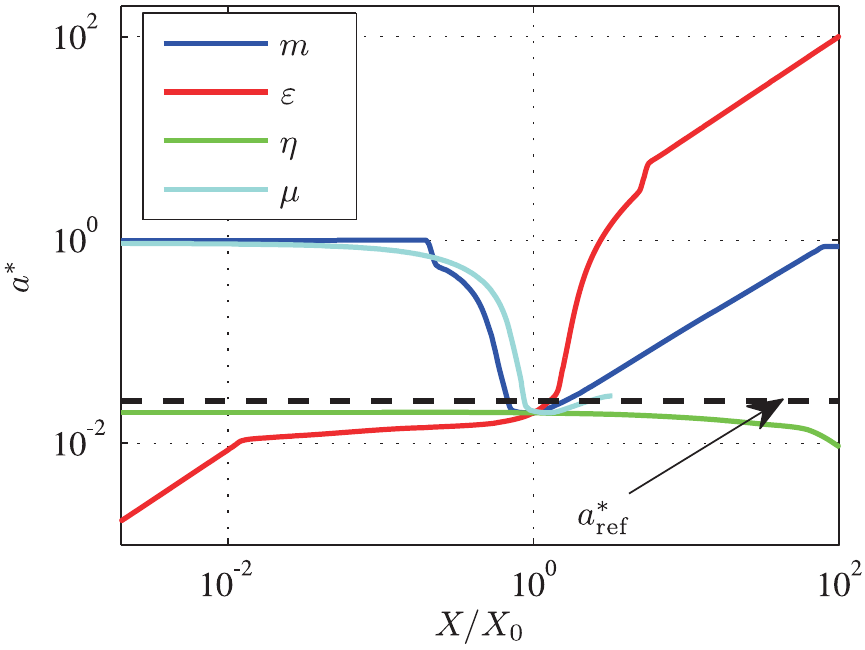}{Sensitivity of maximum forced response amplitude
with respect to uncertain parameters at \g{m=m_0}}{0.7}
\\
The maximum forced response is depicted in \fref{figure08} for
varying system parameters. The influence of the mass corresponds to
the frequency response results illustrated in \fref{figure07}b. In
the field of friction damping, the amplitude-mass characteristic is
also known as optimization or bucket curve. The amplitude-excitation
level characteristic is also known as damper performance curve
\zo{came1990} and exhibits the often reported flat region where the
damper is most effective. The influence of the linear damping factor
is small near the nominal value of \g{\eta_0=0.15\%} since it is
virtually negligible compared to the friction-induced damping of
several percents, see \fref{figure04}b. The effect of the friction
coefficient is qualitatively similar to the influence of the damper
mass.

\subsection{Reliability optimization\label{sec:reliability}}
\paragraph{Influence of the reference performance}
\fss[th]{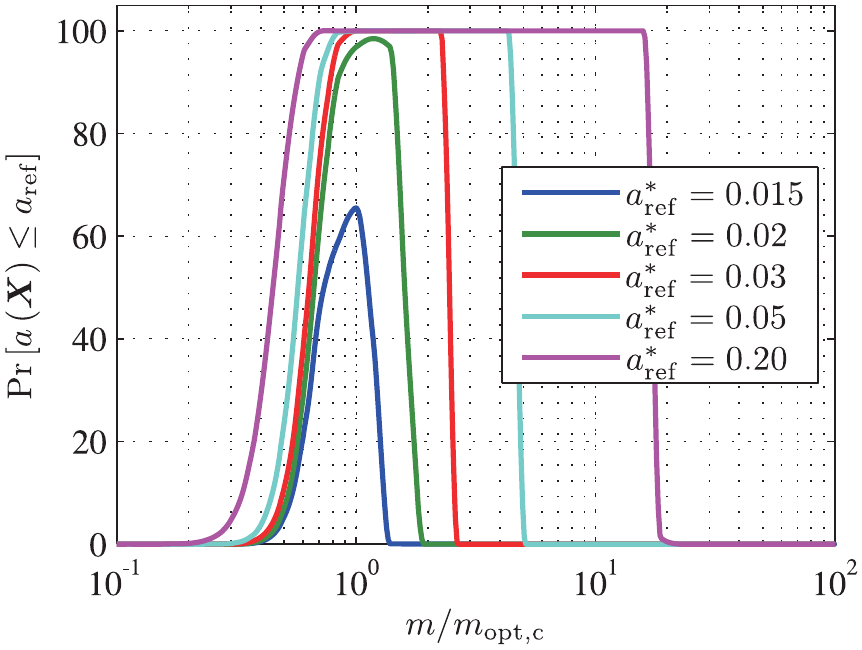}{Reliability for different reference amplitude
values \g{\areference}}{0.7}
The reference performance \g{\areference} has an essential influence
on the reliability characteristic with respect to the design
variable \g{m}, see \fref{figure09}. Only the excitation level was
considered as uncertain in this case with a standard deviation of
\g{15\%}. For very low reference amplitudes, the reliability tends
to zero since a certain amount of forced response cannot be avoided.
For very large reference amplitudes, the reliability tends to
\g{100\%} since the vibration amplitudes will be bounded in general.
Between these extreme cases, the reliability assumes a bell-like
shape. The optimum reliability is typically achieved for mass values
larger than the mass value that minimizes the vibration amplitudes
for nominal parameters. Beyond a certain reference amplitude,
specific ranges of mass values perform equally well with a
reliability of \g{100\%}. It should be noticed that the reference
performance is considered as a parameter that is typically a known
input parameter for a specific design study. For the following
analyses, a value of \g{\areference^*=0.02} is chosen, corresponding
to a desired vibration reduction by a factor of \g{50} compared to
the case without dampers, \cf \fref{figure08}.

\paragraph{Influence of parameter variability}
\fss[th]{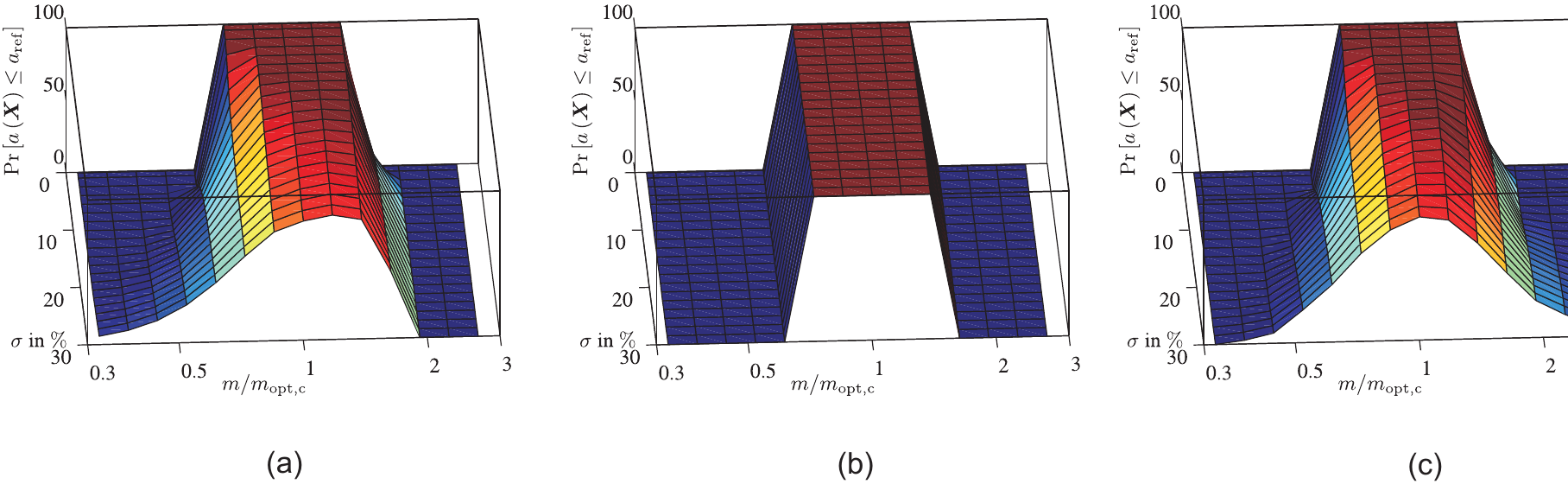}{Reliability as a function of the damper mass and
the variability of the parameters (~(a) excitation level, (b) linear
damping, (c) friction coefficient~)}{1.0}
In \fref{figure10}, the reliability-mass characteristic is
illustrated with respect to the variability of the individual
parameters. For a standard deviation of zero, the reliability is a
step function, \ie it is either one or zero, depending on whether
the reference amplitude is exceeded for nominal parameters. For
increasing variability in the parameters the reliability flattens
with respect to mass. The location of the optimum mass seems not to
be influenced significantly by the variability in the parameters.
The reliability characteristic is a step function for uncertainty in
\g{\eta}, at least up to a standard deviation of \g{30\%}. This is a
consequence of the comparatively low sensitivity of the amplitude
\g{a} with respect to \g{\eta} as illustrated in \fref{figure08}.

\paragraph{Influence of cumulative uncertainty}
\fss[th]{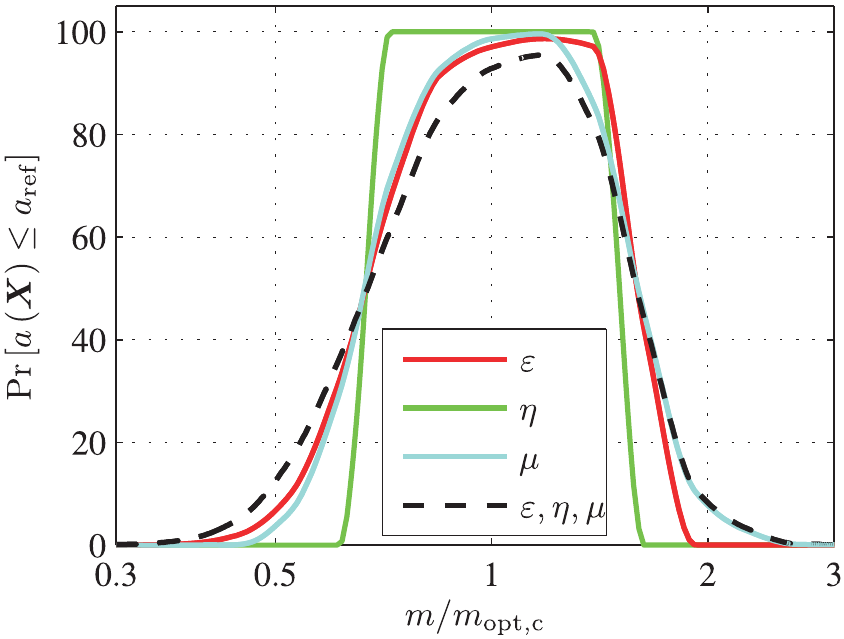}{Reliability for different sets of uncertain
parameters}{0.7}
In \fref{figure11}, the reliability characteristic is illustrated
for individual and cumulative uncertainty in the parameters. The
standard deviation of all parameters was set to \g{15\%}. The
cumulation of uncertainties leads to a flatter shape of the
resulting reliability characteristic compared to the individual
characteristics. The influence is qualitatively similar to the
influence of increasing individual variabilities, see
\fref{figure10}.

\paragraph{Discussion of the optimized damper mass}
As it can be concluded from the previous paragraphs that the optimum
robust design depends on different influences such as the reference
performance and the parameter variabilities. A significant
improvement of the design robustness can only be expected if these
properties are specified with care. For the case when all parameters
are uncertain with a standard deviation of \g{15\%}, the robust
damper mass is \g{17\%} larger than the optimum damper mass without
uncertainties \g{\moptc}, \cf \fref{figure11}. Compared to
\g{\moptc}, the robust mass improves the reliability by \g{2.6\%}
from \g{93\%} to \g{96\%}, but it reduces the nominal performance by
\g{6.7\%} in this case. It should be noticed that these values
clearly depend on the considered system.

\paragraph{Computational cost reduction achieved by the ROM}
\tab[tbh]{c|cccc}{task          & no. of FRF                & comp. time    & comp. time                    & comp. time\\
                                & evaluations               & NMA overhead  & ROM                           & direct (HBM)\\\hline
single FRF                      &   $1$                     &   $1.0$                   &   $0.03$          &   $1.7$\\
\fref{figure09}                 &   $9.2\cdot 10^3$         &   $1.0$                   &   $276$           &   $1.5\cdot 10^4~{}^*$\\
\fref{figure11}                 &   $2.9\cdot 10^6$         &
$10.0$ & $8.9\cdot 10^4$   &   $5.0\cdot 10^6~{}^*$ }{Computational
effort normalized to overhead caused by single NMA, the asterisk in
the last column denotes that the values were estimated based on the
single FRF computation times}{computational_effort}
In the course of computation, it was generally found that there is a
sufficient agreement between ROM and reference calculations. The ROM
is therefore qualified for the reliability optimization.\\
In \tref{computational_effort}, the computation times for different
tasks are listed. It was found that the average computational effort
for computing a single FRF by the direct HBM is in the order of
magnitude of a single NMA computation. This is plausible since the
mathematical problem dimension is almost the same for each solution
point, and a similar number of solution points is required for both
analysis types \zo{krac2013a}. While the direct HBM computes the FRF
branches left and right from the resonance, the NMA determines the
resonant amplitude-frequency relationship, \ie the backbone curve.
In the presented case, the direct HBM computation of a single FRF
takes longer than the NMA computation. It is thus already beneficial
to follow the ROM approach, despite the NMA overhead.\\
The benefit of the ROM approach becomes more prominent as soon as
multiple FRF calculations are required for different parameter sets.
\g{9.2\cdot 10^3} FRF calculation were required to obtain the
results in \fref{figure09}. Since in this case all varied parameters
are readily included in the parameter space of the ROM, the NMA
overhead is not increased compared to a single FRF computation. In
contrast, the friction coefficient is not included in the inherent
ROM parameter space. The NMA was carried out for ten sampling points
of the friction coefficient and interpolation was used between these
points to obtain the results in \fref{figure11}. Hence, the NMA
overhead was larger in this case. Nevertheless, it can be concluded
from \tref{computational_effort} that the NMA overhead is negligible
compared to the actual computations required for the probabilistic
analysis results presented in \fref{figure09} and
\fref{figure11}.\\
Finally, it has to be remarked that the actual computation times
required for the forced response evaluations depend on many
influences such as the problem dimension, type of nonlinearity and
the considered dynamic regime. The number of function evaluations
required for the probabilistic analyses depend on the number of
uncertain parameters and the integration method utilized for the
evaluation of the expectation integral.

\section{Conclusions\label{sec:conclusions}}
A novel robust design approach for friction-damped structures was
proposed and investigated in terms of a proof-of-concept study. The
objective of the design problem is defined as the reliability, \ie
the probability that a specified performance is achieved. For the
case study of a bladed disk with friction dampers, the optimum
design improved the reliability and insensitivity with respect to
uncertain parameters. The robust optimum damper mass was found to be
larger than its counterpart for nominal parameters. It is common
practice to design the damper mass larger than its theoretical
optimum value. The proposed objective gives rise to a probabilistic
justification for this practice and provides means for a
quantitative approaches towards robust design of friction
interfaces for the purpose of vibration damping.\\
The probabilistic nature of the numerical approach results in a
severely increased computational burden compared to the analysis of
the deterministic model. The proposed approach was found to become
prohibitive in conjunction with models of realistic complexity.
Hence, a drastic ROM approach based on the concept of nonlinear
modes was introduced to the probabilistic design approach. A
recently developed multi-harmonic multi-modal ROM was therefore
extended regarding the parameter space required for design approach.
The computational effort was reduced by several orders of
magnitude.\\
Several opportunities for future work arise from the findings of
this paper. It would be desirable to alleviate the simplifications
made for the numerical example and investigate a more realistic case
study. Regarding bladed disks, it would be interesting to
investigate the influence of blade-by-blade varying parameters.

\begin{appendix}
\section{Proof of scale invariance for systems with piecewise
linear contact constraints\label{asec:proof_scaling}} In
\zo{krac2013a}, the scale invariance was postulated in the form of a
hypothesis. The validity was demonstrated by numerical examples only
and no strict mathematical proof was provided. The proof is derived
for the initial value problem in time domain in order to show its
broad range of validity. For convenience, the ODE in \eref{eqm} is
written in state space notation,
\ea{\nonumber \mm{\dot y}(t)+\vector{-\mm{\dot u(t)}\\\mm
M^{-1}\left(\mm K\mm u(t)+\mm C\mm{\dot u}(t)+\fnl\left(\mm
u(t),\mm{\dot u}(t),\preload\right)-\mm f_{\mathrm e}(t)\right)}=\mm
0\fk\\
\mm y(t=0)=\mm y_0\fp}{eqm_statespace}
Herein, the state vector is \g{\mm y^{\mathrm T}=\left[\mm
u^{\mathrm T}\uudist\mm{\dot u}^{\mathrm T}\right]}. The parameter
\g{\preload} can also be a linear function in the components of
displacement \g{\mm u} or velocity \g{\mm{\dot u}}. For this
parameter, the scale invariance in \eref{scaling_property} is
derived in the following.\\
Note that the validity of \eref{scaling_property} is obvious in the
linear case with \g{\fnl\equiv 0}. In the nonlinear case, the
validity of \eref{scaling_property} boils down to the requirement
\e{\fnl\left(\scal\mm u(t),\scal\mm{\dot u}(t),\scal \preload\right)
= \scal\fnl\left(\mm u(t),\mm{\dot
u}(t),\preload\right)\fp}{scalability_nonlinear_force}
The validity of \eref{scalability_nonlinear_force} can now be shown
element-wise for the vector of nonlinear forces \g{\fnl}. In this
study, the proof is restricted to the unilateral spring nonlinearity
and the elastic Coulomb nonlinearity defined in \eref{nttc}, since
these are the only nonlinearities involved in the contact constraint
formulations employed in this study.\\
For the unilateral spring nonlinearity, the parameter \g{\preload}
is the preload \g{p_0}. The requirement in
\eref{scalability_nonlinear_force} can be easily verified by
substitution:
\e{p_{\mathrm n}\left(\scal u_{\mathrm n},\scal p_0\right)
=\left(k_{\mathrm n}\scal u_{\mathrm n}+\scal
p_0\right)_+=\scal\left(k_{\mathrm n} u_{\mathrm n}+
p_0\right)_+=\scal p_{\mathrm n}\left(u_{\mathrm
n},p_0\right)\fp}{n_scaling}
For each contact node, the tangential friction dynamics depend on
the normal dynamics. Since the normal contact law is scalable, also
the limiting friction force \g{\mu p_{\mathrm n}} is scalable. Thus,
for convenience, \g{\preload=\mu p_{\mathrm n}} can be defined as
scaling parameter for the friction law. With this, the requirement
in \eref{scalability_nonlinear_force} can be proven for the friction
law:
\ea{\nonumber\mm p_{\mathrm t}\left(\scal \mm u_{\mathrm
t},\scal \preload^*\right) &=& \begin{cases}\mm 0 & \text{lift-off}\\
k_{\mathrm t}\left(\scal\mm u_{\mathrm t}-\scal\mm u_{\mathrm
t}\left(t_{\mathrm{stick}}\right)\right)+\scal\mm p_{\mathrm
t}\left(t_{\mathrm{stick}}\right) & \text{stick}\\
\scal \preload^*\frac{\scal\mm{\dot u}_{\mathrm
t}}{\|\scal\mm{\dot u}_{\mathrm t}\|} & \text{slip}\end{cases}\\
\nonumber &=& \scal
\begin{cases}\mm 0 & \text{lift-off}\\ k_{\mathrm
t}\left(\mm u_{\mathrm t}-\mm u_{\mathrm
t}\left(t_{\mathrm{stick}}\right)\right)+\mm p_{\mathrm
t}\left(t_{\mathrm{stick}}\right) & \text{stick}\\
 \preload^*\frac{\mm{\dot u}_{\mathrm
t}}{\|\mm{\dot u}_{\mathrm t}\|} & \text{slip}\end{cases}\\
&=& \scal\mm p_{\mathrm t}\left(\mm u_{\mathrm
t},\preload^*\right)\fp}{tc_scaling_1}
The proof needs to be completed by demonstrating the scalability of
the transition conditions:
\ea{\nonumber\text{\underline{transition conditions}:}&\\
\nonumber \text{stick-slip transition: }& \|\scal\mm p_{\mathrm
t}\|=\scal \preload^*\,\,\Leftrightarrow\,\, \|\mm p_{\mathrm
t}\|=\preload^*\fk\\
\text{slip-stick transition: }& \|k_{\mathrm t}\scal\mm{\dot
u}_{\mathrm t}\| = \scal\dot \preload^*\,\,\Leftrightarrow\,\,
\|k_{\mathrm t}\mm{\dot u}_{\mathrm t}\| = \dot \preload^*
\fp}{tc_scaling_2}
Note that the validity of the scale invariance in time domain such
as in \eref{eqm} implies the validity in frequency domain such as in
the NMA. Hence, the scale invariance can be exploited to obtain the
solution of nonlinear eigenproblem for a parameter value \g{\scal
\preload} from the computed solution for \g{\preload}. Thus, the
treatment of the normal preload as an inherent parameter of the ROM,
as proposed in \sref{nmrom} is an exact approach in terms of the
contact constraints used in this study.
\end{appendix}

\end{document}